\documentclass[twocolumn,aps,prc,superscriptaddress]{revtex4-2}

\usepackage{color,amsmath,amssymb,epsfig,graphicx}
\usepackage{bm}
\usepackage{mathptmx, courier, pifont}
\usepackage[scaled=0.92]{helvet}
\usepackage[T1]{fontenc}
\usepackage{textcomp}
\usepackage[colorlinks=true,linkcolor=blue,filecolor=blue,urlcolor=blue,citecolor=blue]{hyperref}

\usepackage{multirow}
\usepackage{longtable}
\UseRawInputEncoding
  \def\ack{\,|\,}

\begin{document}

\title{ Novel chiral-wobbling excitation mode in atomic nuclei }
\author{Kouser Qureshie}
\affiliation{Department of Physics, Islamic University of Science and Technology, Jammu and Kashmir, 192 122, India}
\author{S. P. Rouoof}\email{sprouoofphysics27@gmail.com}
\affiliation{Department of Physics, Islamic University of Science and Technology, Jammu and Kashmir, 192 122, India}
\author{J. A. Sheikh}\email{sjaphysics@gmail.com}
\affiliation{Department of Physics, University of Kashmir, Srinagar,
  Jammu and Kashmir, 190 006, India}
\author{Nazira Nazir}
\affiliation{Institute of Physics (IOP), Sachivalya Marg, Bhubaneswar-751005, India}
\affiliation{Homi Bhabha National Institute, Training School Complex, Anushakti Nagar, Mumbai-400 094, India}
\affiliation{Department of Physics, University of Kashmir, Srinagar,
  Jammu and Kashmir, 190 006, India}
\author{G. H. Bhat}
\affiliation{Department of Physics, Government Degree College Shopian, Jammu and Kashmir Higher Education Department, 192 303, India}
\author{S. Jehangir}
\affiliation{Department of Physics, Government Degree College Kulgam, Jammu and Kashmir Higher Education Department, 192 231, India}
\author{N. A. Rather}
\affiliation{Department of Physics, Islamic University of Science and Technology, Jammu and Kashmir, 192 122, India}
\author{R. J. Guo}
\affiliation{Shandong Provincial Key Laboratory of Nuclear Science, Nuclear Energy Technology and Comprehensive Utilization, School of Nuclear Science, Energy and Power Engineering, Shandong University, Jinan, Shandong 250061, China}
\author{S. Y. Wang}
\affiliation{Shandong Provincial Key Laboratory of Nuclear Science, Nuclear Energy Technology and Comprehensive Utilization, School of Nuclear Science, Energy and Power Engineering, Shandong University, Jinan, Shandong 250061, China}
\author{C. Liu}
\affiliation{Shandong Provincial Key Laboratory of Nuclear Science, Nuclear Energy Technology and Comprehensive Utilization, School of Nuclear Science, Energy and Power Engineering, Shandong University, Jinan, Shandong 250061, China}
\author{Q. B. Chen}
\affiliation{School of Physics, East China Normal University, Shanghai 200241, China}
\author{S. Frauendorf}
\affiliation{Department of Physics, University of Notre Dame, Notre Dame, Indiana 46556,  USA}

\begin{abstract}

  The properties of high-spin band structures in $^{74}$Br nucleus are investigated using the triaxial projected shell model (TPSM)
  and particle triaxial rotor (PTR) approaches.  
  It has already been demonstrated  that lowest 
  three bands observed in $^{74}$Br have a characteristic feature of a chiral-wobbler.
  The first excited band is the chiral
  partner of the yrast band and the second excited band is the wobbling band based on the yrast state. We have analyzed the third excited
  band using the TPSM and PTR calculated quantities, and a unique character
  of this band is revealed. It is predicted that
  this band is the chiral partner of the second excited band, and is also a wobbling mode based on the first excited band. 
     
\end{abstract}

%\keywords{Keywords are listed below the abstract of the manuscript. At least three keywords are required at submission.}

\maketitle

%\section{Introduction}

A microscopic description of quantum many-body systems in terms of the interactions among the constituent particles is one of the
most challenging problems in physics. 
The exact solution of the Schrodinger equation for many-body systems is impossible, except for
those consisting of a few particles, and a mean-field based approach is employed to seek the approximate solution of these systems \cite{RS80}.
The advantage of the mean-field approach is that it incorporates correlations through breaking of the symmetries that the original Hamiltonian
obeys. Although the initial mean-field solution may obey symmetries, but
these can be broken in the iterative process due to the non-linearity inherent in the approximate approach \cite{SF2001,JAS21}. The
solution of the mean-field equations
corresponds to a state which, in general, has broken symmetries. The broken symmetries include rotational, translational, reflection
and gauge symmetry associated with the particle number. This spontaneous symmetry breaking mechanism is associated with phase transitions, giving
rise to the Nambu-Goldstone bosons \cite{Nambu1960,Goldstone1961}.

In molecular and nuclear physics, the breaking of rotational symmetry gives rise to a deformed mean-field solution. The ground-state with
broken rotational symmetry is infinitely degenerate as any arbitrary rotation in space leads to the same energy. In nuclear
physics, the deformed mean-field Nilsson potential has played a pivotal role to unravel the intrinsic structures of deformed nuclei. This 
model has allowed to classify the rich band structures observed in deformed and transitional nuclei \cite{BMII,Nilsson1969}. The majority of nuclei
can be described by using the axially-symmetric
deformed Nilsson potential. There are selection rules based on axial symmetry, for example, the
Alaga rules \cite{PhysRev.100.432}, which
are obeyed by most of the deformed
nuclei in the low-spin region. Nevertheless, there are also a few regions \cite{ ABERG1990c35,PhysRevLett.62.2084,BENGTSSON198956,Bengtsson2004}
in the periodic table that are predicted to have broken axial symmetry, which are referred to
as the triaxial nuclei. The triaxial nuclei have drawn a considerable attention in recent years with the experimental realization
of chiral symmetry and wobbling mode in some regions of the Segre chart. These two phenomena are only possible for triaxial nuclei.
The chiral doublet bands are obtained by rotating about an axis which does not lie in a principal-plane and is, therefore, known as an aplanar rotation.
The three mutually perpendicular angular momenta, formed by short- intermediate- and long- axis, can be arranged to form two systems with opposite
chirality, namely left- and right-handed systems. They are transformed into each other by the chiral operator that combines time-reversal and a spatial
rotation by $180^\circ$. The criteria for the observation of chiral rotation are: (i) near degeneracy of the excitation energy spectra of the doublet bands,
(ii) the energy staggering parameter should be a smooth function of angular momentum, (iii) similar moments of inertia and $B(E2)$
transitions for the two bands, and (iv)
staggering in the $B(M1)$ transitions with phase of in-band transitions opposite to that of the inter-band transitions.
The chiral doublet bands
were predicted in 1997 \cite{FM97} and were first observed in 2001 \cite{PhysRevLett.86.971}. Since then these bands 
have been identified in several mass
regions \cite{XIONG2019193,PhysRevLett.86.971,PhysRevC.63.051302,PhysRevC.63.061304,PhysRevC.64.031304,PhysRevLett.91.132501,PhysRevLett.99.172501}.

The wobbling motion is another phenomenon that is possible only for a triaxial density distribution in which the rotational axis
does not coincide with any of the principal axes. It is described as the harmonic oscillation of one of the three
principal axes of a triaxial rotor
about the space-fixed total angular momentum vector. The precession of a free asymmetric top in classical mechanics is
analogue of the wobbling mode
in quantum physics. The fingerprints of the wobbling motion are: (i) almost equally spaced band structures with alternating
even- and odd-spin states, (ii) inter-band $B(E2)$ transitions are comparable to the in-band transitions, and (iii) $B(M1)$ transitions are very weak.
The wobbling mode was first predicted by Bohr and Mottelson for even-even systems \cite{BMII}. However, it was first experimentally observed
in the odd-system of $^{163}$Lu \cite{OHJ2001}. This mode has now been reported in several
other nuclei \cite{WM44,WM45,WM2,WM46,WM54,WM47,WM48,WM6,WM12}.

The above chiral doublet and wobbling band structures have been observed for different nuclides. However, in a more recent investigation,
both these modes have been proposed to coexist in the odd-odd $^{74}$Br nucleus \cite{Guo24}. This is the first candidate of the
coexistence of chiral and wobbling modes
in atomic nuclei. It has been shown that the first and second excited bands are, respectively, the chiral partner
and the wobbling mode based on the yrast band. This inference has been drawn from the analysis of the experimental data, and also from the
calculations performed using the particle triaxial rotor (PTR) model \cite{Guo24}.

The purpose of the present work is to investigate the chiral-wobbling mode observed in $^{74}$Br using the microscopic approach
of triaxial projected shell model (TPSM). It is demonstrated from the analysis of TPSM calculated quantities that the lowest
three bands observed in $^{74}$Br can be characterized as originating from the chiral-wobbling mode as proposed in the earlier
investigation \cite{Guo24}. We have also evaluated the properties of the fourth lowest band and it is
predicted that this band is the chiral partner of the third band, and is also the wobbling excitation mode based on the second band. The
novel results of the present work are the following $:$ the first excited band is a chiral partner of the yrast band, the second excited band is the wobbling band based on the yrast band, and the third excited band is the chiral partner
of the second excited band and is also the wobbling mode built on the first excited band. We have also performed the PTR model
calculations, and it is shown that the fourth band has the features as obtained in the TPSM analysis.

The TPSM approach employed in the present work is now an established microscopic tool to describe the high-spin properties in
deformed and transitional nuclei \cite{JG12, GH14, SJ18, ishfaq20, JS21b, JS21, nazira22, nazira, Rouoof2024}.
In several studies \cite{WM39, NaziraNP}, it has been demonstrated to describe the properties of the chiral doublet bands quite accurately.
Recently, the TPSM approach has been employed to provide a microscopic description of the wobbling mode in atomic nuclei \cite{SJ24ch, bhat2025}. In a detailed investigation,
the wobbling mode observed in odd-mass nuclei has been studied and it has been shown that TPSM provides an accurate description of
the measured properties. In particular, the coexistence of wobbling  and the signature partner bands observed in several nuclei
has been corroborated in the TPSM approach \cite{JG12,WM39,SJ24ch,bhat2025}. Recently, wobbling mode has been studied for even-even systems of
$^{76}$Ge, $^{112}$Ru, $^{188,192}$Os, $^{192}$Pt and $^{232}$Th and it has
been demonstrated that $\gamma$-bands observed in these nuclei, except for $^{232}$Th, develop into wobbling structures after the
quasiparticle alignment \cite{bhat2025}. This has been established following the analysis of the excitation energies and the electromagnetic
transition probabilities. 

The TPSM approach is based on the pairing plus quadrupole-quadrupole Hamiltonian of Kumar and Baranger \cite{Kumar1968}.
In this approach,
the triaxial Nilsson potential is 
employed as the mean-field potential \cite{SH99} instead of performing the Hartree-Fock-Bogoliubov (HFB) transformation as
Nilsson potential is the mean-field of the quadrupole-quadrupole interaction. The pairing interaction is treated using the Bardeen-Cooper-Schrieffer (BCS) approximation \cite{RS80}. The resulting Nilsson+BCS
wavefunctions are then projected onto good angular momentum states using the three-dimensional angular momentum projection
operator \cite{KY95}.

To describe the high-spin states, the TPSM
basis space contains multi-quasiparticle configurations. For instance, to describe the high-spin properties of odd-odd systems, the basis
states are composed of one-proton plus one-neutron, three-proton plus one-neutron, and one-proton plus three-neutron quasiparticle configurations.
The projected basis from these multi-quasiparticle configurations are then employed to diagonalize
the shell model Hamiltonian
consisting of pairing plus quadrupole-quadrupole interaction terms, which is expressed as 
\begin{equation}
\hat H = \hat H_0 - {1 \over 2} \chi \sum_\mu \hat Q^\dagger_\mu
\hat Q^{}_\mu - G_M \hat P^\dagger \hat P - G_Q \sum_\mu \hat
P^\dagger_\mu\hat P^{}_\mu ,
\label{hamham}
\end{equation}
where $\hat H_0$ in the above equation is the spherical single-particle potential \cite{Ni69}.
The pairing parameter $G_M$ is chosen such that the
observed odd-even mass differences are
reproduced, and the quadrupole pairing strength
$G_Q$ is assumed to be 0.18 times $G_M$.
The QQ-force
strength $\chi$ is obtained by the self-consistency 
relation, \mbox{$2 \hbar\omega~\epsilon= 3\chi \langle\Phi\ack\hat Q_0\ack\Phi\rangle$} \cite{KY95}.

\begin{figure}[!ht]
    \begin{center}
        \hspace*{-0.2cm}\includegraphics[width=8.5cm, angle =0]{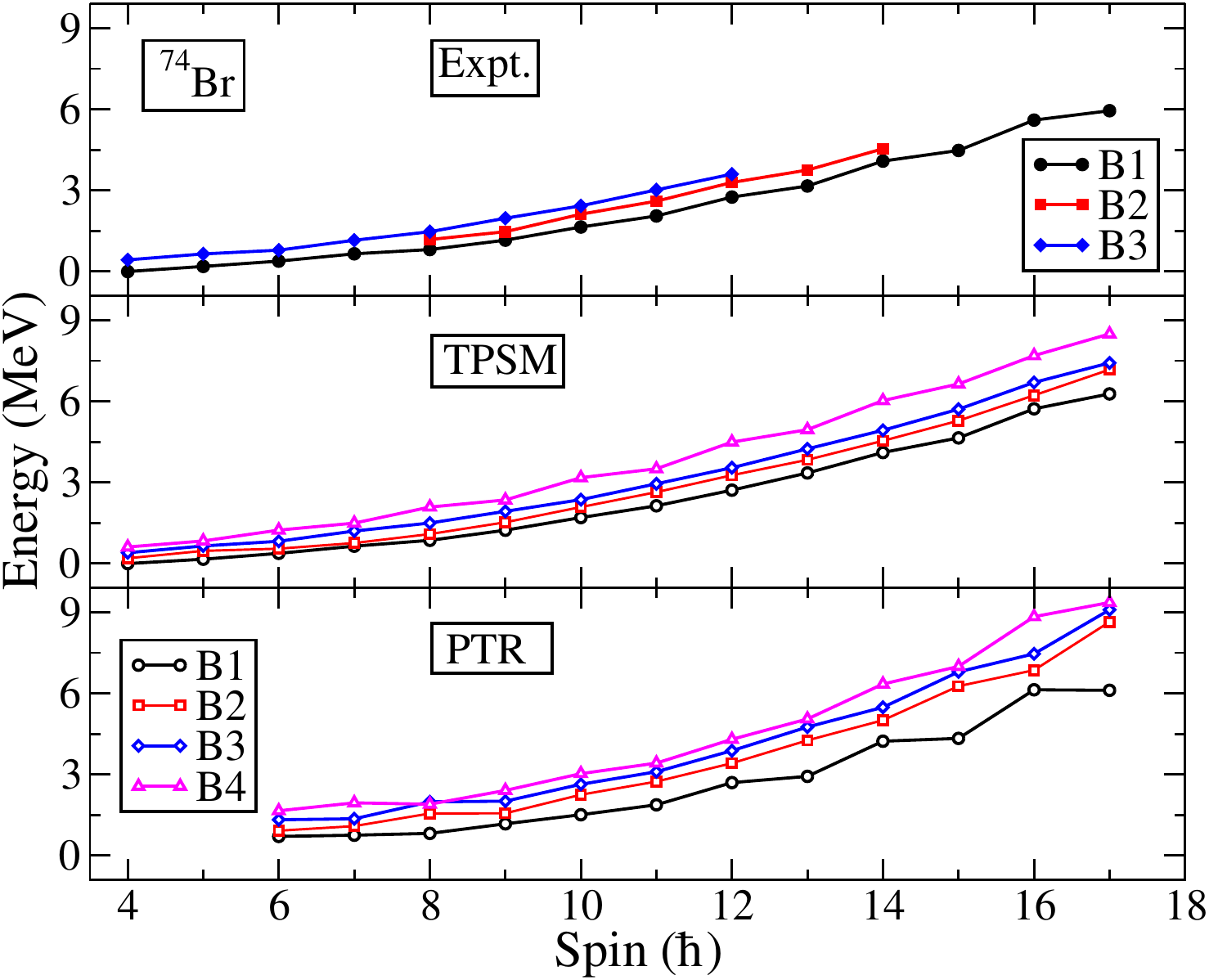}
    \end{center}
    \vspace{-0.5cm}
    \caption{Experimental and calculated excitation energies for the lowest four bands, labelled as B1, B2, B3, and B4 in $^{74}$Br.}
  
    \label{fig1}
\end{figure}

The TPSM calculations have been performed for $^{74}$Br with the deformation parameters of $\beta=0.367$ and $\gamma=27^\circ$ \cite{MOLLER1995185}, which are similar to those used in the PTR investigation with $\beta=0.45$ and $\gamma=27.5^\circ$ \cite{Guo24}.
The results of the excitation energies, obtained from the TPSM calculations, are compared with the corresponding
experimental quantities in Fig.~\ref{fig1}. In the figure, the energies obtained from the PTR model for the lowest four bands
are also displayed in the bottom panel. The PTR results have been obtained with the same parameter set as used in Ref.~\cite{Guo24}.
It is evident from the figure that the TPSM and PTR approaches reproduce the experimental energies for the
known states fairly well. The bands B1 and B2 have been interpreted as the chiral partner bands, although the
energy differences are somewhat larger than expected. In the ideal chiral limit, the two bands should have identical excitation energies. 
The energy differences in the proposed chiral bands in Fig.~\ref{fig1} are similar to other chiral doublet bands
identified for the odd-odd Bromine isotopes \cite{WANG201140,PhysRevLett.116.112501,PhysRevC.100.054309,XU2022137287}.

%===========fig1========================================
%======================================================
%===========fig2========================================
\begin{figure}[!ht]
    \begin{center}
        \hspace*{-0.2cm}\includegraphics[width=8.5cm, angle =0]{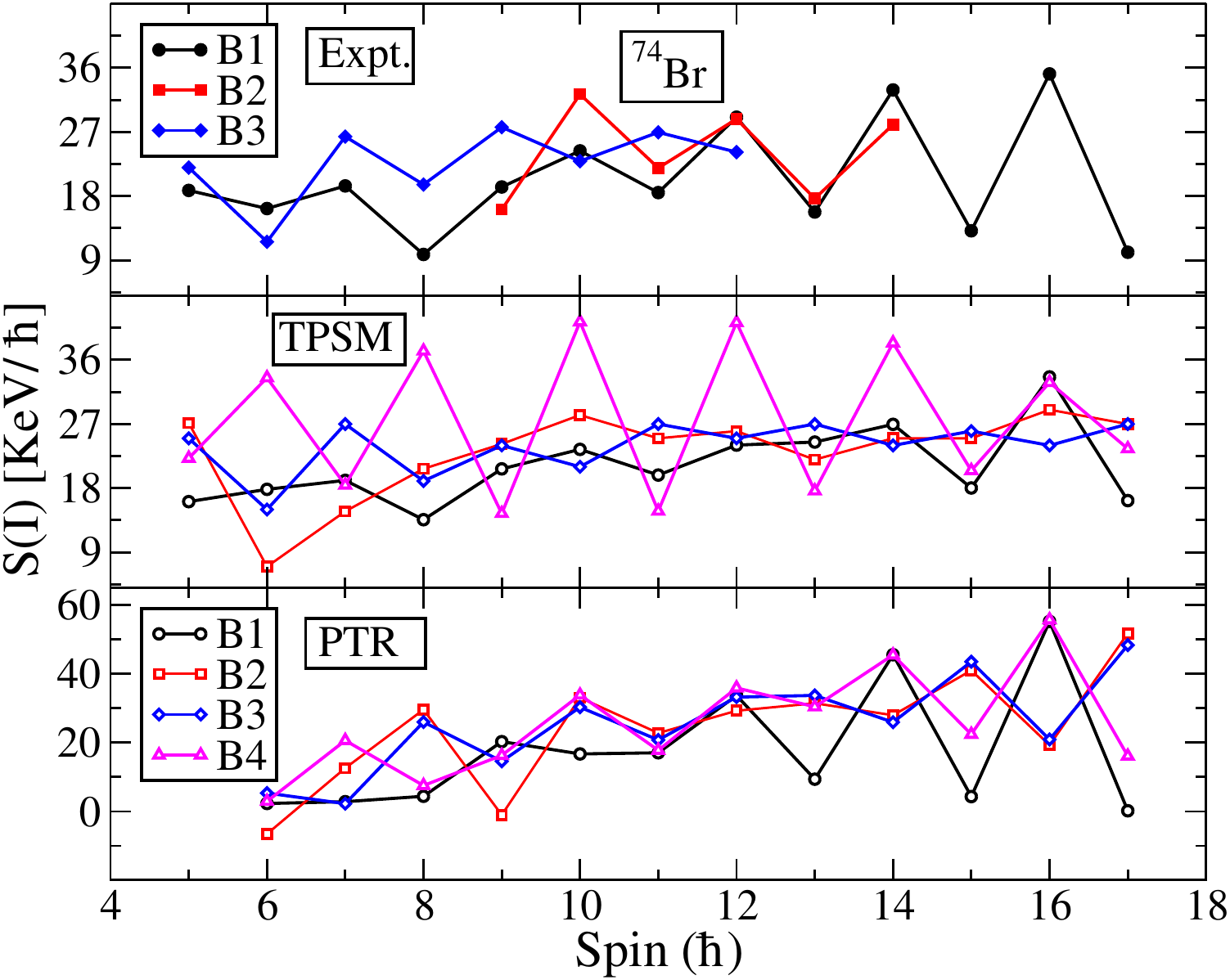}
    \end{center}
    \vspace{-0.5cm}
    \caption{Experimental and calculated excitation energy staggering parameters $S(I)=[E(I)-E(I-1)]/2I$ for the lowest four bands, labelled as B1, B2, B3, and B4 in $^{74}$Br.}
  
    \label{fig2}
\end{figure}
%======================================================
%%===========fig3========================================
\begin{figure}[!ht]
    \begin{center}
        \hspace*{-0.2cm}\includegraphics[width=8.5cm, angle =0]{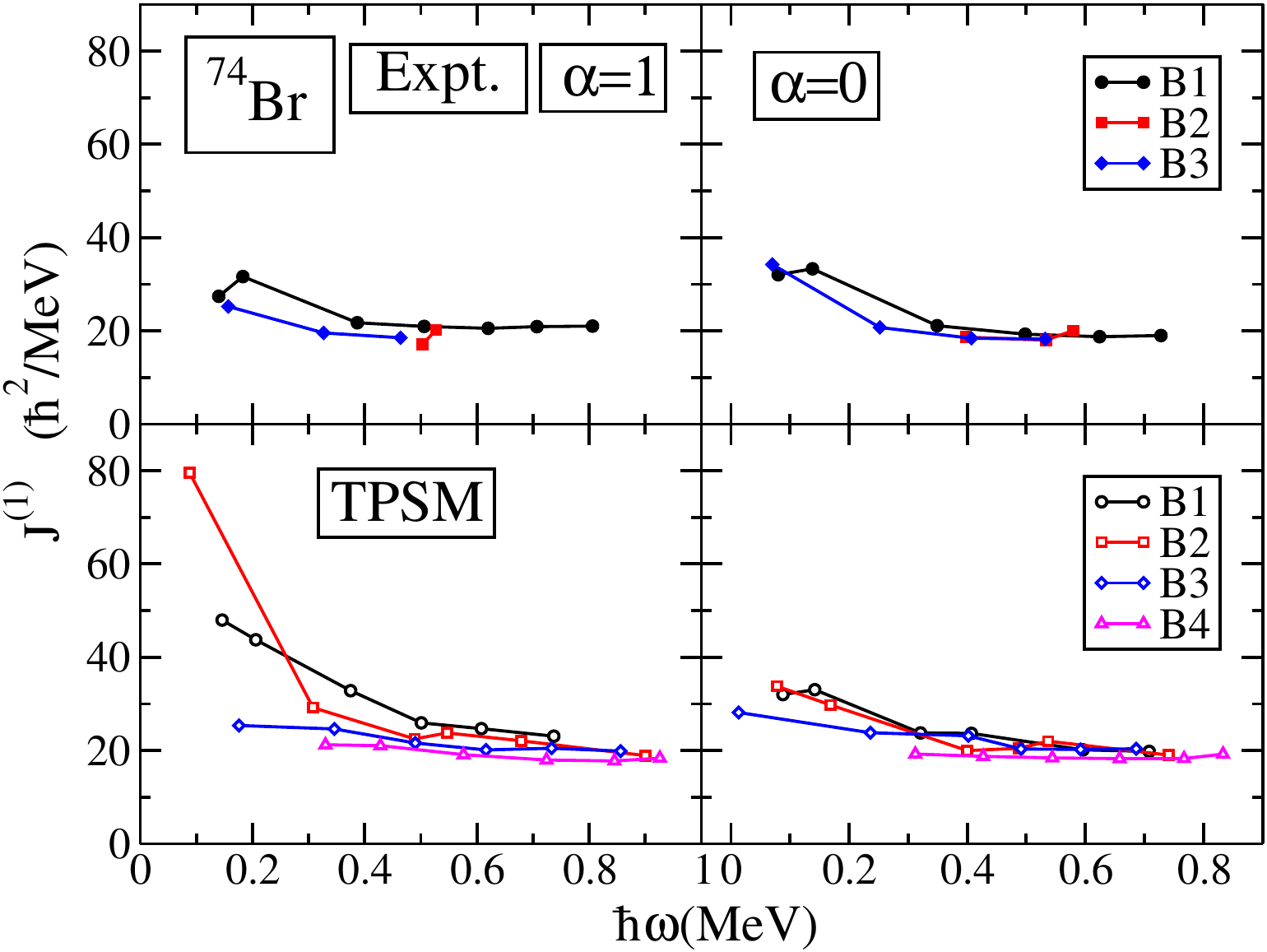}
    \end{center}
    \vspace{-0.5cm}
    \caption{ Kinematic moments of inertia $J^{(1)}$ of the lowest four bands, labelled as B1, B2, B3, and B4 in $^{74}$Br.
  }
    \label{fig5}
\end{figure}
%=======================================================
%===========fig4========================================
\begin{figure}[!ht]
    \begin{center}
        \hspace*{-0.2cm}\includegraphics[width=8.5cm, angle =0]{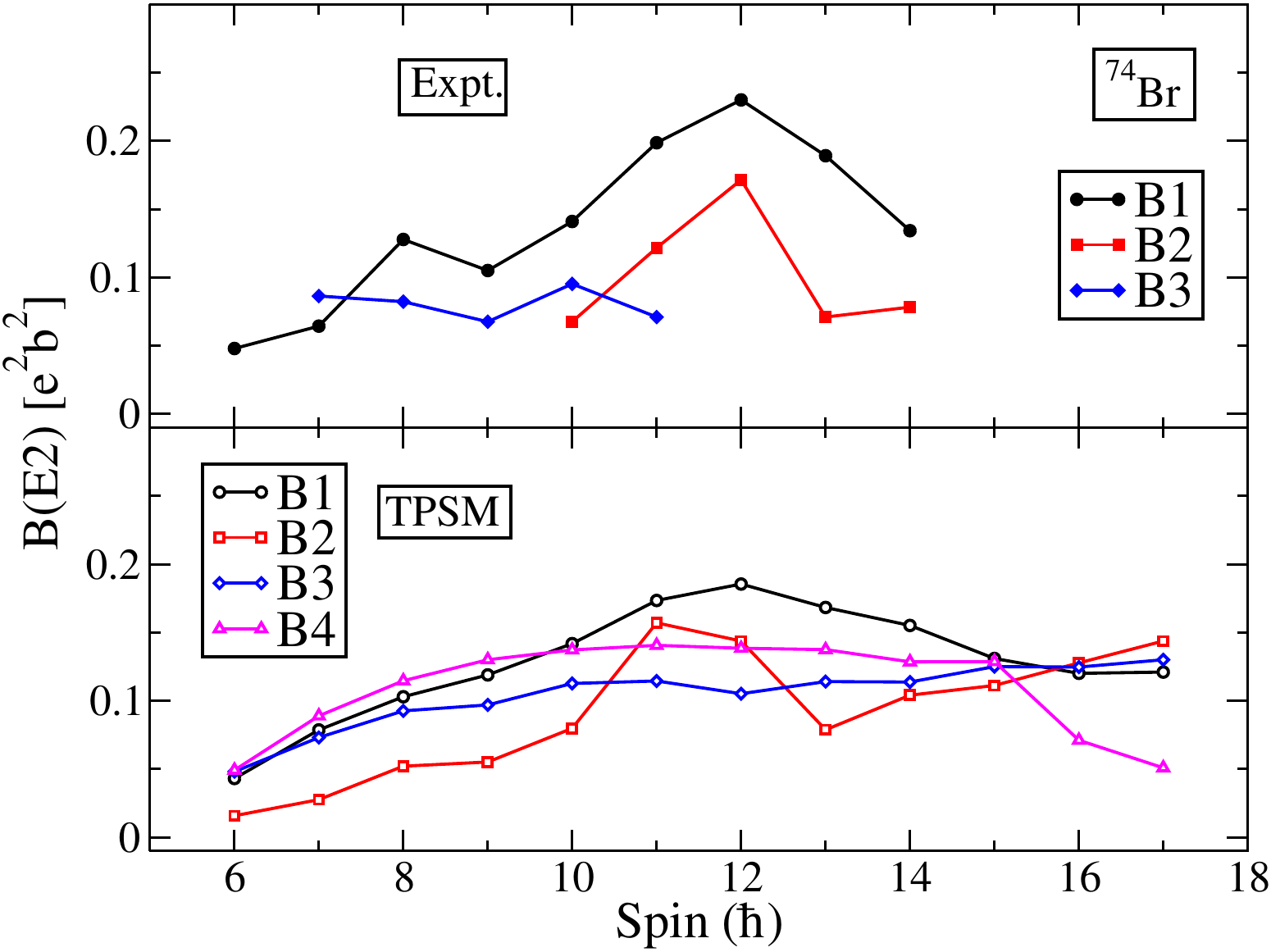}
    \end{center}
    \vspace{-0.5cm}
    \caption{Experimental and calculated absolute in-band transition probabilities $B(E2)$ for the lowest four bands, labelled as B1, B2, B3, and B4 in $^{74}$Br.
  }
    \label{fig3}
\end{figure}
%===========fig5=======================================

\begin{figure}[!ht]
    \begin{center}
        \hspace*{-0.2cm}\includegraphics[width=8.5cm, angle =0]{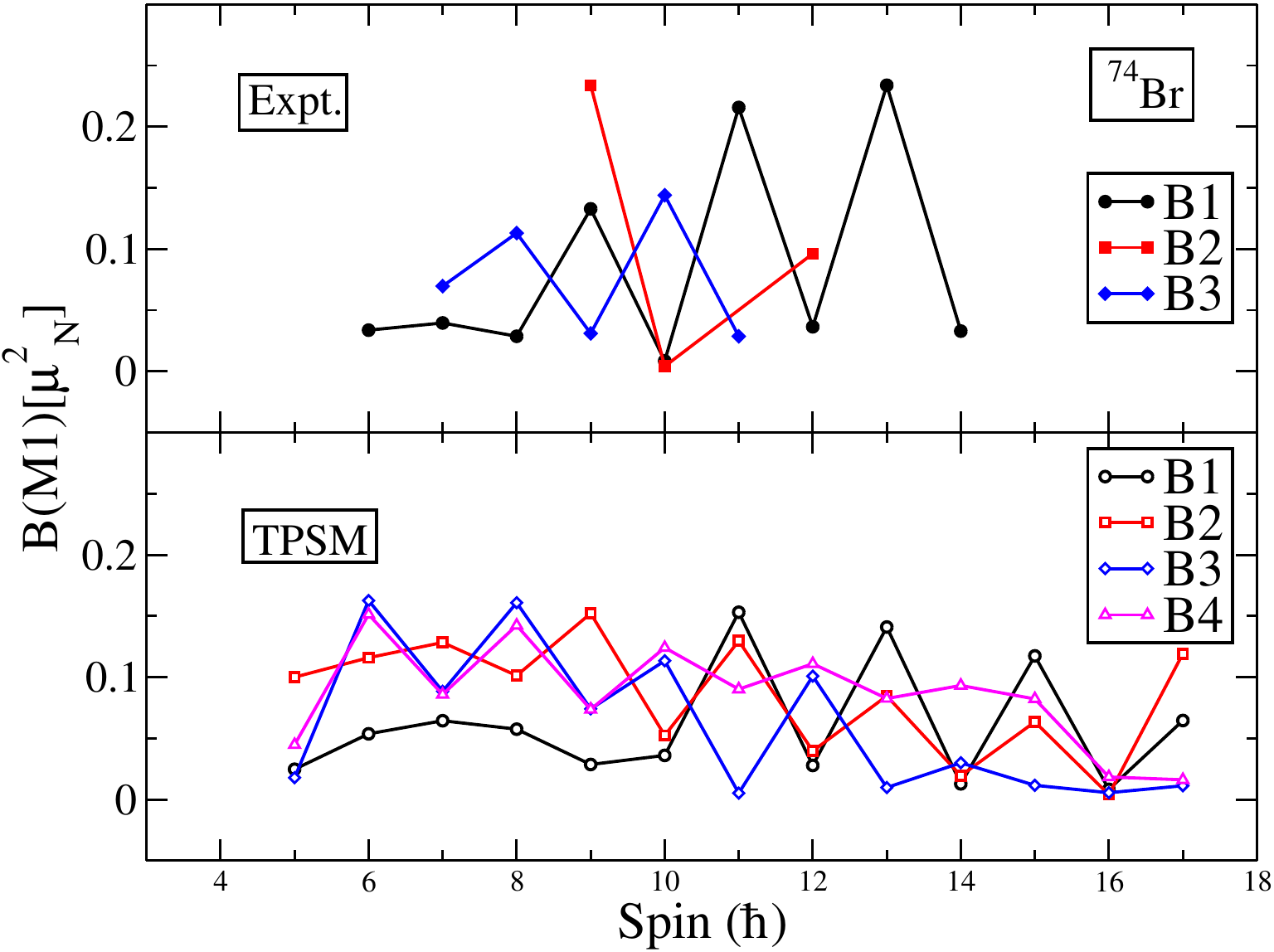}
    \end{center}
    \vspace{-0.5cm}
    \caption{ Experimental and calculated absolute in-band transition probabilities $B(M1)$ for the lowest four bands, labelled as B1, B2, B3, and B4 in $^{74}$Br.  }
    \label{fig6}
\end{figure}
%=============================================================================================================
%===========fig6========================================
\begin{figure}[!ht]
    \begin{center}
        \hspace*{-0.2cm}\includegraphics[width=8.5cm, angle =0]{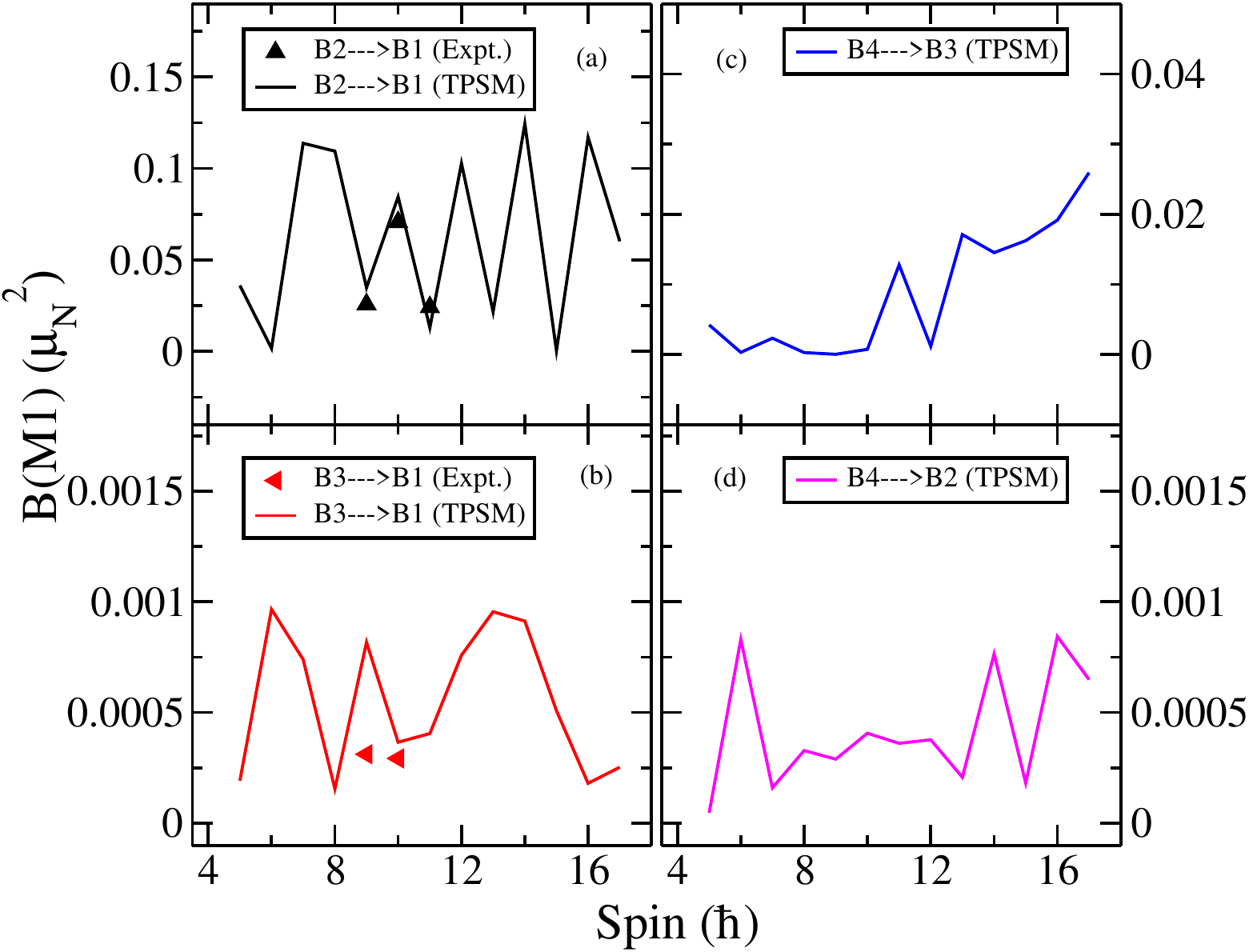}
    \end{center}
    \vspace{-0.5cm}
    \caption{Experimental and calculated $B(M1)$ of inter-band $\Delta I=1$ transitions between bands B1 and B2, B1 and B3, B2 and B4 and B3 and B4. }
    \label{fig7}
\end{figure}
%======================================================
%===========fig7========================================
\begin{figure}[!ht]
    \begin{center}
        \hspace*{-0.2cm}\includegraphics[width=8.5cm, angle =0]{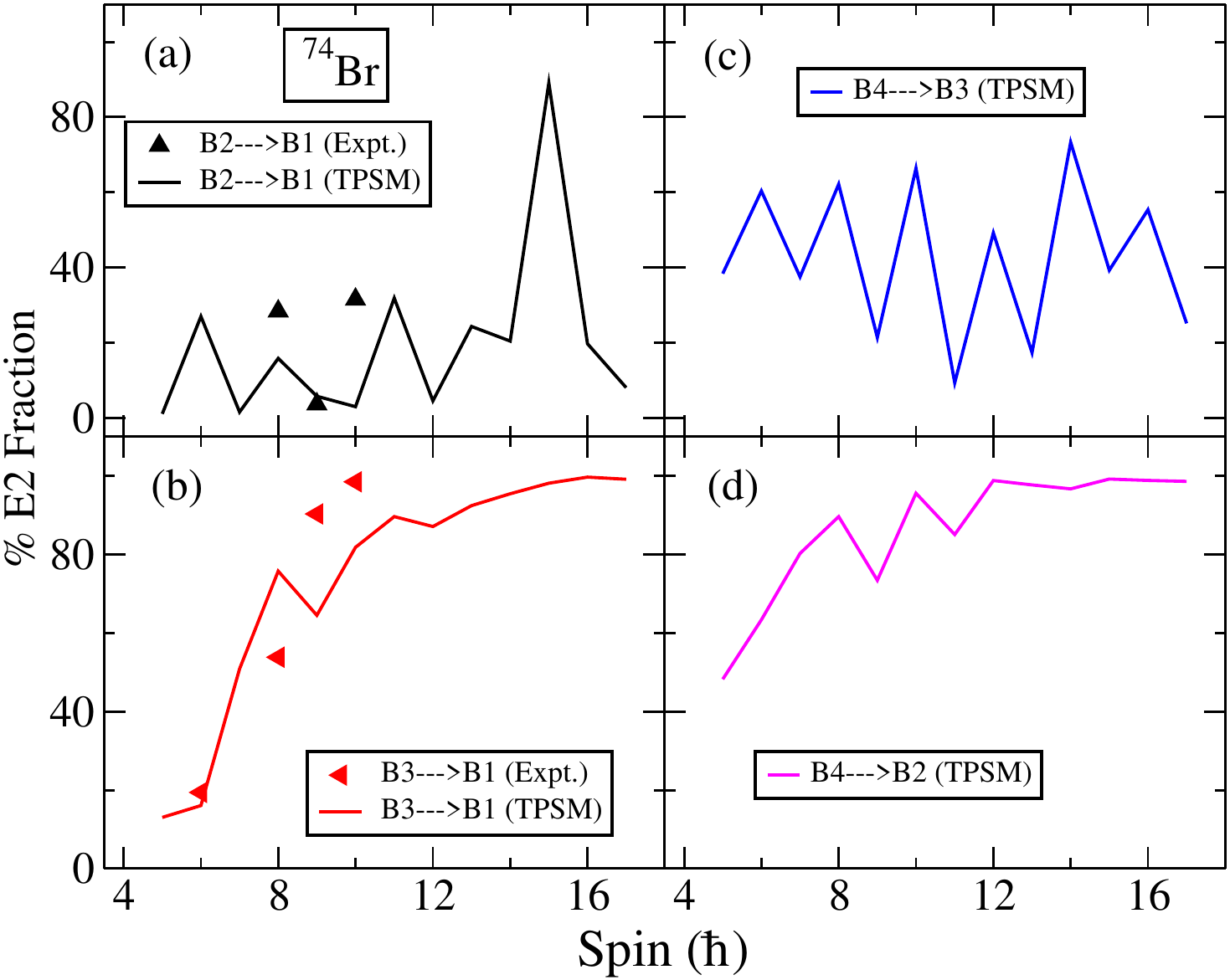}
    \end{center}
    \vspace{-0.5cm}
    \caption{Experimental and calculated $E2$ fraction of inter-band $\Delta I=1$ transitions between bands B1 and B2, B1 and B3, B2 and B4 and B3 and B4.   }
    \label{fig8}
\end{figure}
%======================================================
%===========fig8========================================
\begin{figure}[!ht]
    \begin{center}
        \hspace*{-0.2cm}\includegraphics[width=8.5cm, angle =0]{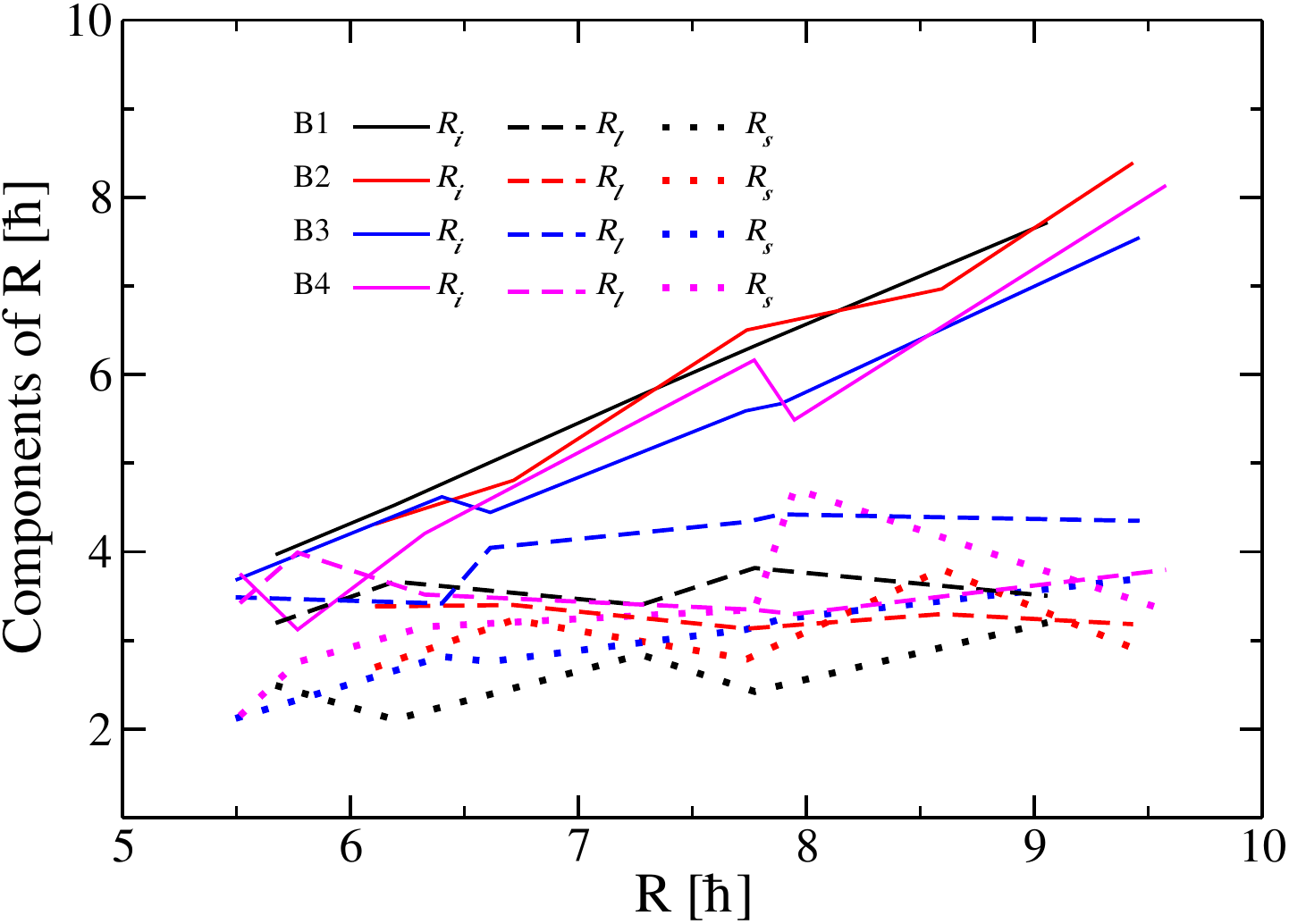}
    \end{center}
    \vspace{-0.5cm}
    \caption{Root-mean-square angular momentum components
along the intermediate (i), long (l), and short (s) axes of the
core
angular momentum ${\bf R}$ as a function of $R$ $\biggl(\sqrt{R_i^2+R_s^2+R_l^2}\biggr)$ for bands B1 (black lines), B2 (red lines), B3
(blue lines), and B4 (magenta lines). }
    \label{fig9}
\end{figure}
%======================================================

The even- and odd-spin members of the observed band B3 in Fig.~\ref{fig1} have been suggested as wobbling structures based on
odd- and even-spin members of the band B1, respectively. In the figure, we have also included the energies calculated
using TPSM and PTR approaches for the third excited band, labelled as B4, and it is
evident from the figure that this band has similar energies as those of other bands. It will be
demonstrated in the following that this band is the chiral partner of band B3, and is also a wobbling excitation mode built on band B2.

To explore the chiral and wobbling features of the four bands, we shall now analyze the energy staggering, 
moment of inertia and electromagnetic 
properties of these bands. The energy staggering parameter, $S(I)$, is depicted in Fig.~\ref{fig2} for the three experimental bands, and TPSM
and PTR calculated four bands. For chiral bands, the energy staggering parameter should have a weak dependence on spin. In Fig.~\ref{fig2},
$S(I)$ shows some amount of staggering with spin, but is similar to other Bromine isotopes where chiral symmetry has been proposed to
exist \cite{Guo24}.
The calculated moments of inertia, $J^{(1)}$, from the experimental data and the TPSM approach are displayed in Fig.~\ref{fig5}. It is evident
from the figure that these are similar for all the bands as expected for the chiral and wobbling band structures.

The transition probabilities have been evaluated using the TPSM wavefunction with the effective charges of $e_n=0.5, e_p=1.5$
for $B(E2)$ transitions, and $g_l=0(1.0)$ and $g_s=-3.826(5.58)$ for neutron (proton) g-factors for $B(M1)$ transitions. The
calculated $B(E2)$ in-band transitions are depicted
in Fig.~\ref{fig3} along with the known experimental values. It is noted from the figure that these transitions are similar for the
four band structures. The $B(E2)$ transitions are expected to be similar for chiral and wobbling band structures as they are based
on the same intrinsic configuration. 

A distinct feature of the chiral bands is provided by the phase of the $B(M1)$ transitions. It has been shown that in the
chiral limit \cite{hamamoto172502}, the phase of the in-band $B(M1)$ transitions is opposite to that of the inter-band transitions. These transitions
are displayed in Figs.~\ref{fig6} and \ref{fig7} along with the observed values for the known transitions. It is noted from the figures that
the phase of the in-band transitions for bands B1 and B2 is same, and the inter-band transitions between the two bands
have opposite phase. This phase correlation is one of the important characteristic property of the chiral bands. In an
analogous manner, the bands B3 and B4 have a similar phase relationship with the in-band transitions having
same phase, but the inter-band transitions bear the opposite phase. Therefore, bands B3 and B4 are predicted to be chiral
band structures. It is observed from Fig.~\ref{fig6} that in-band transitions of B3 have opposite phase
to those of band B1, and has been interpreted as wobbling structure based on band B1. The interesting inference drawn
from Fig.~\ref{fig6} is that
band B4 has also opposite phase to band B2. Hence, it is predicted that band B4 is the wobbling mode based on band B2.

To substantiate further the above chiral-wobbling mode, the $\%E2$ fraction for the band structures is displayed in Fig.~\ref{fig8}. For
chiral bands, the $M1$ transitions
dominate and this quantity should be small, whereas for the wobbling mode, $E2$ transitions dominate and the $\%E2$ fraction should be quite large.
It is noted from Fig.~\ref{fig8} that the $\%E2$ fraction is small for B2 to B1 transitions and, therefore, these transitions are dominated by $M1$. The $\%E2$
fraction for the transitions from band B3 to B1 is quite large, indicating that the transitions are dominated by $E2$.  In a similar manner, the transitions
from band B4 to B3 are dominated by $M1$, whereas the transitions from B4 to B2 are dominated by $E2$. These observations on the inter-band transitions
suggest the presence of a unique chiral-wobbling excitation mode in $^{74}$Br.

From geometrical considerations, the difference between chiral and wobbling motion is the contribution
of the collective rotor angular momentum. For chiral bands, the rotor contributions are similar for the doublet bands, and
in the case of wobbling bands, the rotor angular momentum of the $n_{\omega}=1$ band should be lower than that of the $n_{\omega}=0$
band, since the rotational axis for the excited wobbling band moves away from the principal axis.
The collective rotor angular momentum contributions of the four bands along the three axes \cite{Guo24} have been evaluated
using the PTR wavefunctions, and are
displayed in Fig.~\ref{fig9}. In comparison to the TPSM approach, the rotor and the particle angular momenta are separated in the PTR approach. It is evident from
the figure that intermediate, i-axis is the rotational axis as rotor angular momentum has maximum contribution along this
axis, and the contributions from short- and long- axis are small. It is noted that bands B1 and B2 have similar rotor contributions and,
in comparison, bands B3 and B4 have about one unit lower values. Further, bands B3 and B4 have similar contributions for
high-spin states. These observations suggest that pairs of bands, (B1, B2) and (B3, B4) are chiral bands, and
B3 (B4) is the wobbling band based on B1 (B2).

In conclusion, it has been shown  using the TPSM and PTR approaches that $^{74}$Br has a unique chiral-wobbling
excitation mode in atomic nuclei. The first excited band is the chiral partner of the yrast band, the second excited band is
the wobbling band based on the yrast band, and the third excited band is the chiral partner of the second excited band and is
also the wobbling mode based on the first excited band. The characteristic features of the lowest three bands have already been
established in the experimental work, and it is of considerable interest to confirm the predictions of the present
work for the fourth band. Further, it is expected that the unique chiral-wobbling mode predicted for $^{74}$Br
may also be present in other nuclides, and we are presently investigating other Br-isotopes and also other regions to explore
the possibility of observing this excitation mode.

\bibliographystyle{apsrev4-2}
\bibliography{CW}

\end{document}